\documentclass[12pt]{article}
\usepackage{amssymb}
\usepackage{amstext}
\usepackage{amsmath}
\usepackage{latexsym}
\usepackage{color}

\newcommand{\be}{\begin{equation}}
\newcommand{\en}{\end{equation}}

\newtheorem{thm}{Theorem}
\newtheorem{cor}[thm]{Corollary}

\newtheorem{defi}{Definition}[section]
\newtheorem{lem}[defi]{Lemma}
\newtheorem{theorem}[defi]{Theorem}

\newcommand{\bedefin}{\begin{defi}}
\newcommand{\findefi}{\end{defi} \medskip}

\newcommand{\betheo}{\begin{theorem}$\!\!${\bf \,\,\,}}
\newcommand{\entheo}{\end{theorem}}
\newcommand{\enth}{\end{theorem}}

\newcommand{\becor}{\begin{cor}$\!\!${\bf .}}
\newcommand{\encor}{\end{cor}}

\newcommand{\belem}{\begin{lem}$\!\!${\bf .}}
\newcommand{\enlem}{\end{lem}}

\newcommand{\bea}{\begin{eqnarray}}
\newcommand{\ena}{\end{eqnarray}}

\newcommand{\beano}{\begin{eqnarray*}}
\newcommand{\enano}{\end{eqnarray*}}

\newcommand{\bee}{\begin{enumerate}}
\newcommand{\ene}{\end{enumerate}}

\newcommand{\bei}{\begin{itemize}}
\newcommand{\eni}{\end{itemize}}

\newcommand{\betab}{\begin{tabular}}
\newcommand{\entab}{\end{tabular}}

\newcommand{\bd}{\begin{displaymath}}

\newcommand{\h}{{\mathfrak H}}

\newcommand{\htil}{\widetilde{\mathfrak H}}

\newcommand{\lh}{{\mathcal L}({\mathfrak H})}

\newcommand{\hs}{\mathcal B_2 (\h )}

\newcommand{\bfeta}{\mbox{\boldmath $\eta$}}

\newcommand{\bPhi}{\mbox{\boldmath $\Phi$}}
\newcommand{\bPsi}{\mbox{\boldmath $\Psi$}}

\newcommand{\bcalX}{\mbox{\boldmath $\mathcal X$}}

\newcommand{\bcalH}{\mbox{\boldmath $\mathcal H$}}

\newcommand{\bH}{\mathbf H}

\newcommand{\bP}{\mathbf P}

\catcode `\@=11 \@addtoreset{equation}{section}

\catcode `\@=12

\begin{document}
\baselineskip 18pt plus 2pt minus 2pt

\begin{center}
{\LARGE \bf  Modular Structures on Trace Class \\
        Operators and Applications\\[5mm]
         to Landau Levels}\\[10mm]

{\large S. Twareque Ali} \footnote[1]{Department of Mathematics and
Statistics, Concordia University,
Montr\'eal, Qu\'ebec, CANADA H3G 1M8\\
e-mail: stali@mathstat.concordia.ca}
\vspace{3mm}\\

{\large F. Bagarello} \footnote[2]{ Dipartimento di Metodi e Modelli
Matematici,
Facolt\`a di Ingegneria, Universit\`a di Palermo, I-90128  Palermo, ITALY\\
e-mail: bagarell@unipa.it\,\,\,\, Home page:
www.unipa.it/$^\sim$bagarell}
\vspace{3mm}\\

{\large G. Honnouvo}\footnote[3]{Department of Mathematics and Statistics,
McGill University, Burnside 1021, 805 Sherbrooke W.
Montreal, Qu\`ebec,
CANADA H3A 2K6}
\end{center}

%--------------------------------------------------------------

\begin{abstract}
\baselineskip 18pt plus 2pt minus 2pt
The energy levels, generally known as the Landau levels, which characterize the motion of
an electron in a constant magnetic field, are those of the one-dimensional harmonic oscillator,
with each level being infinitely degenerate. We show in this paper how the associated
von Neumann algebra of observables display a modular structure in the sense of the
Tomita-Takesaki theory, with the algebra and its commutant referring to the two orientations
of the magnetic field. A KMS state can be built which in fact is the Gibbs state for an
ensemble of harmonic oscillators. Mathematically, the modular structure is shown to arise
as the natural modular structure associated to the Hilbert space of all Hilbert-Schmidt
operators.

\end{abstract}

\newpage

%---------------------------------------------------------------

\section{Introduction}\label{sec:intro}
The motion of an electron in a constant electromagnetic field is a well known problem
in atomic physics. Quantum mechanically, the energy levels of such a system, which are
generally known as the Landau levels (see, for example, \cite{cohtan}), are linearly
spaced, with each level being infinitely degenerate. Indeed, the energy levels are
exactly those of the harmonic oscillator, with infinite degeneracy at each level. The
Hamiltonian of the system can be written as the sum of two oscillator Hamiltonians, together
with an interaction part, which is an angular momentum term. It turns out that the
diagonalized Hamiltonian resembles that of a single harmonic oscillator, with
infinite muliplicity at each level. If the sense of the magnetic field is reversed,
one obtains a second Hamiltonian, similar to the first, but commuting with it. Both these
Hamiltonians can be sritten in terms of two pairs of mutually commuting oscillator type
creation and annihilation operators, which then
generate two von Neumann algebras which mutually commute, and in fact are commutants
of each other. This leads to the existence of a modular
structure, in the sense of the Tomita-Takesaki theory \cite{takesaki}. The invariant state
of the theory turns out to be the Gibbs state for an ensemble of harmonic oscillators,
the modular operator, giving the time-evolution under which this state is invariant, is directly
obtained from the interaction Hamiltonian and the modular conjugation operator simply
interchanges the two possible orientations of the magnetic field.  Preliminary discussions
of some aspects of the theory presented in this paper have been discussed in
\cite{aag_book,alibag,aliemch} and  \cite{emch}. However, here we present a unified discussion,
along with a physical interpretation and explore holomorphic aspects of the theory, its
connection to families of orthogonal polynomials (Hermite and complex Hermite) and to
various related families of coherent states.

  The rest of the paper is organized as follows. In Section \ref{sec:summ-math} we briefly
recall the main features of the Tomita-Takesaki modular theory of von Neumann algebras;
in Section \ref{sec:setup} we work out a simple example of this theory in the space of
Hilbert-Schmidt operators on a Hilbert space; in Sections \ref{sec:applications},
\ref{sec:second-rep} and  \ref{sec-phys-disc} we give a detailed analysis of the
problem of the electron in a constant magnetic field in the light of the modular theory,
bringing out the physical meaning of the its various mathematical ingredients. In
Section \ref{sec:diag-CS} we look at some associated families of coherent states.
Finally in Section \ref{sec:concl} we give our final comments.  Certain
mathematical properties of von Neumann algebras, which are required in the paper, are
collected in the Appendix.

\section{Summary of the  mathematical theory}\label{sec:summ-math}

This Section is devoted to a quick review of the Tomita-Takesaki modular theory of
von Neumann algebras, to the extent that it it is needed in this paper.
Details and proofs of statements may
be found, for example,  in \cite{sewell,stratila,takesaki}. Some basic definitions
and  notions about  von Neumann algebras are listed in the Appendix. Let $\mathfrak A$ be a von
Neumann algebra on a Hilbert space $\h$ and $\mathfrak A^\prime$ its commutant. Let
$\bPhi \in \h$ be a unit vector which is cyclic and separating for $\mathfrak A$.  Then
the corresponding state $\varphi$ on the algebra, $\langle \varphi\; ; \; A\rangle =
\langle \bPhi \mid A \bPhi \rangle\; , \; A \in \mathfrak A\; , $ is faithful and normal.
Consider the antilinear map,
\be
  S:\h \longmapsto \h\; , \qquad SA\bPhi = A^*\bPhi\; , \; \forall A \in \mathfrak A \; .
\label{mod-antilin-map}
\en
Since $\bPhi$ is cyclic, this map is densely defined and in fact it can be shown that it is
closable. We denote its closure again by $S$ and  write its polar decomposition as
\be
  S = J\Delta^\frac 12 = \Delta^{-\frac 12}J\; , \quad \text{with} \quad \Delta = S^*S\; .
\label{mod-pol-decomp}
\en
The operator $\Delta$, called the {\em modular operator}, is positive and self-adjoint. The
operator $J$, called the {\em modular conjugation operator}, is antiunitary and satisfies
$J = J^*\; , \; J^2 = I_\h$. Note that the antiunitarity of $J$ implies that
$\langle J\phi \mid J\psi \rangle = \langle \psi \mid \phi\rangle\; , \; \forall \phi, \psi
\in \h$.

  Since $\Delta$ is self-adjoint, using its spectral representation, we see that for $t\in
\mathbb R$, the family of operators $\Delta^{-\frac {it}\beta}$, for some fixed $\beta > 0$,
defines a unitary family of automorphisms of the algebra $\mathfrak A$. Denoting these
automorphisms by $\alpha_\varphi (t)$, we may write,
\be
  \alpha_\varphi (t)[A] = \Delta^{\frac {it}\beta}A\Delta^{-\frac {it}\beta}\; , \;\;
  \forall A \in \mathfrak A\; .
\label{mod-automorph-grp}
\en
Thus, they constitute a strongly continuous one-parameter group of automorphisms, called the
{\em modular automorphism group\/}.
Denoting the generator of this one-parameter group by $\mathbf H_\varphi$, we
get
\be
   \Delta^{-\frac {it}\beta} = e^{it\mathbf H_\varphi} \quad \text{and} \quad
     \Delta = e^{-\beta \mathbf H_\varphi}\; .
\label{mod-automorph-grp2}
\en
It can then be shown that the state $\varphi$ is invariant under this automorphism group:
\be
  e^{-\beta \mathbf H_\varphi}\bPhi = \bPhi\; ,
\qquad  \Delta^{\frac {it}\beta}\;\mathfrak A \;
    \Delta^{-\frac {it}\beta} = \mathfrak A\; ,
\en
and the antilinear map $J$ interchanges $\mathfrak A$ with its commutant $\mathfrak A^\prime$:
\be
  J\mathfrak A J = \mathfrak A^\prime\; .
\label{mod-interch}
\en

  Finally, the state $\varphi$ can be shown to satisfy the {\em KMS (Kubo-Martin-Schwinger)
condition\/}, with respect to the automorphism group $\alpha_\varphi (t)\; , \; t \in
\mathbf R$, in the following sense. For any two $A,B \in \mathfrak A$, the  function
\be
   F_{A, B} (t) = \langle\varphi\; ;\; A\; \alpha_\varphi (t)
   [B]\rangle\; ,
\label{mod-KMS-func}
\en
has an extension to the strip $\{ z = t+ iy\mid t \in \mathbb R,\; y \in [0, \beta]\} \subset
\mathbb C$ such that $F_{A, B} (z)$ it is analytic in the open strip $(0, \beta )$ and
continuous on its boundaries. Moreover, it also satisfies the boundary condition ({\em at inverse
temperature} $\beta$),
\be
  \langle\varphi\; ;\; A\; \alpha_\varphi (t + i\beta ) [B]\rangle =
   \langle\varphi\; ;\; \alpha_\varphi (t) [B]  \;A\rangle \; , \quad t \in \mathbb R\; .
\label{mod-KMS-prop}
\en

\section{A simple example of the theory}\label{sec:setup}

A simple example of the Tomita-Takesaki theory and its
related KMS states can be built on the space of Hilbert-Schmidt operators on a Hilbert space.
The set of Hilbert-Schmidt operators is itself a Hilbert space, and there are two
preferred algebras of operators on it, which carry the modular structure. The presentation
here follows that in \cite{aag_book} (Chapter 8, Section 4). A detailed application of
this structure to Landau levels is discussed in Section \ref{sec:applications}, which
extends some recent work reported in \cite{alibag}.

Again,  $\h$ be a (complex, separable) Hilbert space of dimension $N$  (finite or infinite) and
$\{\zeta_i\}_{i=1}^N$  an orthonormal basis of it ($\langle \zeta_i \mid \zeta_j \rangle =
\delta_{ij}$). We denote by $\mathcal B_2 (\h ) \simeq \h \otimes \overline{\h}$ the
space of all Hilbert-Schmidt operators on $\h$. This is a Hilbert space with scalar
product
$$ \langle X\mid Y\rangle_2 = \text{Tr}[X^* Y]\; . $$
The vectors,
\be \{X_{ij} = \vert\zeta_i\rangle\langle \zeta_j \vert \mid i,j = 1,2,\ldots , N\}\; ,
\label{basis-vects}
\en
form an orthonormal basis of  $\mathcal B_2 (\h ) $,
$$ \langle X_{ij}\mid X_{k\ell}\rangle_2  = \delta_{ik}\delta_{\ell j}\; . $$
In particular, the vectors,
\be
    \mathbb P_i = X_{ii} = \vert\zeta_i \rangle\langle \zeta_i \vert \; ,
\label{proj-op1}
\en
are one dimensional projection operators on $\h$. In what follows
 $I$ will denote the identity operator on $\h$ and  $I_2$ that on $\hs$.

   We identify a special class of linear operators on $\mathcal B_2 (\h ) $,
denoted by $A\vee B , \; A,B \in \mathcal L (\h)$, which act on a vector $X \in
\mathcal B_2 (\h ) $ in the manner:
$$    (A\vee B) (X) = AXB^*\; . $$
Using the scalar product in $\hs$, we see that
$$ \rm{Tr} [X^* (AYB^*) ] = \rm{Tr} [(A^*X B)^*  Y) ] \Longrightarrow
  (A\vee B)^* = A^* \vee B^* \; , $$
  and since for any $X \in \hs$
  $$ (A_1\vee B_1) (A_2 \vee B_2) (X) = A_1 [(A_2 \vee B_2 )(X)]B_1^*
    = A_1 A_2 X B_2^* B_1^* \; , $$
we have,
  \be
  (A_1\vee B_1) (A_2 \vee B_2) = (A_1A_2 ) \vee (B_1B_2) \; .
\label{comp-law}
\en

  There are two special von Neumann algebras which can be built out of these operators.
These are,
\be
   \mathfrak A_\ell = \{ A_\ell = A \vee I \mid A \in \mathcal L (\h)\} \; , \qquad
   \mathfrak A_{\rm{r}} = \{ A_{\rm{r}} = I \vee A \mid A \in \mathcal L (\h)\} \; .
\label{left-right-alg}
\en
They are mutual commutants and both are factors:
\be
  (\mathfrak A_\ell)^\prime =  \mathfrak A_{\rm{r}} \; , \qquad
  ( \mathfrak A_{\rm{r}})^\prime  =  \mathfrak A_\ell\; , \qquad
  \mathfrak A_\ell \cap \mathfrak A_{\rm{r}} = \mathbb C I_2\; .
\label{factors}
\en

Consider now the operator $J: \hs \longrightarrow \hs$, whose action on the vectors
$X_{ij}$ in (\ref{basis-vects}) is given by
\be
   JX_{ij} = X_{ji} \Longrightarrow J^2 = I_2 \quad \text{and} \quad J(\vert \phi
   \rangle\langle \psi \vert )  = \vert\psi \rangle\langle \phi\vert\;,
   \quad \forall \phi , \psi \in \h\; .
\label{J-op}
\en
This operator is antiunitary, and since
$$ [J(A\vee I) J]X_{ij} = J (A\vee I)X_{ji} = J(AX_{ji}) = J(A\vert\zeta_j\rangle\langle
\zeta_i\vert = \vert \zeta_i\rangle \langle \zeta_j\vert A^*  = (I \vee A )X_{ij} \; , $$
we immediately get
\be
  J \mathfrak A_\ell J  =  \mathfrak A_{\rm{r}}\; .
\label{J-op2}
\en

\subsection{A KMS state}\label{sec:KMS}
Let $\alpha_i \; , \;\; i=1,2, \ldots , N$ be a sequence of non-zero, positive
numbers, satisfying,
$\sum_{i=1}^N \alpha_i = 1$. Let
\be
  \bPhi =  \sum_{i=1}^N \alpha_i^{\frac 12}\; \mathbb P_i =
   \sum_{i=1}^N \alpha_i^{\frac 12}\; X_{ii} \in  \hs\; .
\label{KMS-vect}
\en
We note the following properties of $\bPhi$.
\medskip
\begin{itemize}
\item[(1)] $\bPhi$ defines a vector state $\varphi$ on the von Neumann algebra
$\mathfrak A_\ell$. This follows from the fact that for any $A\vee I \in \mathfrak A_\ell$,
we may define the state $\varphi$ on $\mathfrak A_\ell$ by
\be
 \langle \varphi \; ; \; A\vee I \rangle  =
 \langle \bPhi \mid  (A\vee I)( \bPhi) \rangle_2  =
     \text{Tr} [\bPhi^* A \bPhi ] = \text{Tr}[\rho_\varphi A ] \; , \quad \text{with}\quad
     \rho_\varphi = \sum_{i=1}^N\alpha_i \mathbb P_i \; ,
\label{vect-state}
\en

\item[(2)] The state $\varphi$ is faithful and normal. Normality follows from the last
equality in (\ref{vect-state}) and the fact that $\rho_\varphi$ is a density matrix.
To check for
faithfulness, note that for any $A\vee I \in \mathfrak A_\ell$,
$$  \langle \varphi \; ; \; (A\vee I)^*(A\vee I) \rangle  = \text{Tr}[\rho_\varphi A^*A]
  = \sum_{i=1}^N \alpha_i \Vert A\zeta_i\Vert^2\; , $$
from which it follows that $ \langle \varphi \; ; \; (A\vee I)^*(A\vee I) \rangle = 0$
if and only
if $A = 0$ (since the $\zeta_i$ are an orthonormal  basis set and the $\alpha_i >0$),
hence if and only if $A\vee I = 0$.

\item[(3)] The vector $\bPhi$ is cyclic and separating for $\mathfrak A_\ell$. Indeed,
cyclicity follows from the fact that if $X \in \hs$ is orthogonal to all
$(A\vee I)\bPhi\; , \;
A \in \lh$, then
$$ \text{Tr} [X^* A\bPhi ] = \sum_{i=1}^N \alpha_i^{\frac 12}\langle\zeta_i \mid X^*
A\zeta_i\rangle = 0 , \qquad \forall A \in \lh\; .$$
Taking $A = X_{k\ell}$,  we easily get from the above equality,
$\langle \zeta_\ell \mid X^* \zeta_k\rangle  = 0$ and since this holds for all
$k, \ell$, we get
$X= 0$. In the same way, $\bPhi$ is also cyclic for $\mathfrak A_{\rm{r}}$, hence separating
for $\mathfrak A_\ell$, i.e., $(A\vee I)\bPhi = (B\vee I)\bPhi \Longleftrightarrow
A\vee I  = B\vee I$.
\end{itemize}

We shall show in the sequel that the state $\varphi$ constructed above is indeed a KMS state for a particular choice of $\alpha_i$.

\subsection{Time evolution and modular automorphism}\label{sec:time-evol}
We now construct a time evolution $\alpha_\varphi (t), \; t \in \mathbb R$, on the algebra
$\mathfrak A_\ell$, using the state $\varphi$, with respect to which it has the KMS property,
for fixed $\beta >0$,
\be
  \langle\varphi\; ;\; A_\ell\; \alpha_\varphi (t + i\beta ) [B_\ell]\rangle =
   \langle\varphi\; ;\; \alpha_\varphi (t) [B_\ell]  \;A_\ell\rangle \; , \quad \forall
   A_\ell, B_\ell \in  \mathfrak A_\ell\; ,
\label{KMS-prop}
\en
and moreover the function,
\be
   F_{A_\ell, B_\ell} (z) = \langle\varphi\; ;\; A_\ell\; \alpha_\varphi (z)
   [B_\ell]\rangle\; ,
\label{KMS-func}
\en
is analytic in the strip $\{\Im (z) \in (0, \beta)\}$ and continuous on its boundaries. We
start by defining the operators,
\be
  \bP_{ij} = \mathbb P_i \vee \mathbb P_j \; , \qquad i,j = 1,2, \ldots , N\;
\label{proj op2}
\en
where the $\mathbb P_i$ are the projection operators on $\h$ defined in (\ref{proj-op1}).
Clearly,
the $\bP_{ij}$ are projection operators on the Hilbert space $\hs$.

   Using $\rho _\varphi$ in (\ref{vect-state}) and for a fixed $\beta > 0$, define the
   operator
 $H_\varphi$ as:
 \be
  \rho_\varphi = e^{-\beta H_\varphi} \Longrightarrow H_\varphi =
  -\frac 1\beta \sum_{i=1}^N \log \alpha_i \mathbb P_i\; .
 \label{hamiltonian1}
 \en
 Next we define the operators:
 \be
  H_\varphi^\ell = H_\varphi \vee I\; , \qquad  H_\varphi^{\rm{r}} =
  I\vee H_\varphi \; , \qquad
  \bH_\varphi =  H_\varphi^\ell - H_\varphi^{\rm{r}} \; ,
 \label{hamiltonian2}
 \en
 Since $\sum_{i=1}^N \mathbb P_i = I$, we may also write
$$
  H_\varphi^\ell = -\frac 1\beta \sum_{i,j =1}^N \log \alpha_i \bP_{ij}\; , \quad
\text{and} \quad
H_\varphi^{\rm{r}} = -\frac 1\beta \sum_{i,j =1}^N \log \alpha_j \bP_{ij}\; . $$
Thus,
\be
 \bH_\varphi =  -\frac 1\beta \sum_{i,j =1}^N \log \left[\frac {\alpha_i}{\alpha_j}\right ]
    \bP_{ij}\; .
\label{hamilton3}
\en
Using the operator.
\be
   \Delta_\varphi : = \sum_{i,j =1}^N \left[\frac {\alpha_i}{\alpha_j}\right ] \bP_{ij}
= e^{-\beta \bH_\varphi}\; ,
\label{delta-op}
\en
we define a time evolution operator on $\hs$:
\be
 e^{i\bH_\varphi t } = [\Delta_\varphi ]^{-\frac {it}\beta}\; . \qquad t \in \mathbb R\; ,
\label{timeev-op}
\en
and we note that, for any $X\in \hs$,
\beano
 e^{i\bH_\varphi t } (X) & = & \sum_{i,j =1}^N
\left[\frac {\alpha_i}{\alpha_j}\right ]^{-\frac {it}\beta} \bP_{ij}(X) =
\left[\sum_{i=1}^N(\alpha_i)^{-\frac {it}\beta}\mathbb P_i\right]\vee
 \left[\sum_{j=1}^N(\alpha_j)^{-\frac {it}\beta}\mathbb P_j (X)\right]\\
 & = &  e^{iH_\varphi t } (X) e^{-iH_\varphi t }\; ,
\enano
so that
\be
  e^{i\bH_\varphi t } =  e^{iH_\varphi t } \vee e^{iH_\varphi t }\; ,
\label{timeev-op2}
\en
where $H_\varphi$ is the operator defined in (\ref{hamiltonian1}). From the
definition of the vector $\bPhi$ in (\ref{KMS-vect}), it is clear that it commutes with
$H_\varphi$ and hence that it is invariant  under this time evolution:
\be
 e^{i\bH_\varphi t } (\bPhi)  = e^{iH_\varphi t }\; \bPhi \; e^{-iH_\varphi t }
  = \bPhi \; .
\label{inv-vect}
\en

  Finally, using $e^{i\bH_\varphi t }$ we define the time evolution $\alpha_\varphi$
on the algebra  $\mathfrak A_\ell$, in  the manner (see (\ref{mod-automorph-grp})):
\be
  \alpha_\varphi (t) [A_\ell ] = e^{i\bH_\varphi t }\; A_\ell \; e^{-i\bH_\varphi t } \qquad
 \forall A_\ell  \in \mathfrak A_\ell \; .
\label{timeev-op3}
\en
Writing $A_\ell = A\vee I\; , \;\; A \in \lh$, and  using the composition law (\ref{comp-law}),
we see that
\be
 e^{i\bH_\varphi t }\; A_\ell \; e^{-i\bH_\varphi t } =
\left[e^{iH_\varphi t }\; A \; e^{-iH_\varphi t }\right]\vee I\; ,
\label{timeev-op4}
\en
so that by virtue of (\ref{vect-state}),
\be
 \langle \varphi\; ; \;  \alpha_\varphi (t) [A_\ell ]\rangle =  \text{Tr}\left[ \rho_\varphi \;
   e^{iH_\varphi t }\; A \; e^{-iH_\varphi t }\right] =
   \langle \varphi\; ; \;  A_\ell \rangle\; ,
\label{inv-state}
\en
since $\rho_\varphi$ and $H_\varphi$ commute. Thus, the state $\varphi$ is invariant
under the time evolution $\alpha_\varphi$.

  To obtain the KMS condition (\ref{KMS-prop}), combining  (\ref{timeev-op3}) and
(\ref{timeev-op4}), we first note that, with  $A_\ell = A\vee I\; , \;\; B_\ell = B\vee I$,
$$ A_\ell \alpha_\varphi (t)[B_\ell] =
\left[ Ae^{iH_\varphi t }\; B \; e^{-iH_\varphi t }\right]
\vee I \; .$$
Hence, again using (\ref{vect-state}),
$$ F_{A_\ell , B_\ell} (t) = \langle \varphi \; ; \; A_\ell\; \alpha_\varphi (t)[B_\ell]\rangle
= \text{Tr} \left[ \rho_\varphi A e^{iH_\varphi t }\; B \; e^{-iH_\varphi t }\right]
= \text{Tr} \left[ \rho_\varphi e^{-iH_\varphi t }\; A \; e^{iH_\varphi t } B\right]\; , $$
the last equality following from the commutativity of $\rho_\varphi$ and $H_\varphi$.
Thus, since $\rho_\varphi = e^{-\beta H_\varphi}$,
$$ F_{A_\ell , B_\ell} (t + i\beta ) =
\text{Tr} \left[ \rho_\varphi e^{-iH_\varphi t }e^{\beta H_\varphi}\; A \; e^{iH_\varphi t }
e^{-\beta H_\varphi} B\right] =\text{Tr} \left[e^{-iH_\varphi t }\; A \; e^{iH_\varphi t }
\rho_\varphi  B\right] \; , $$
so that
$$\langle \varphi \; ; \; A_\ell\; \alpha_\varphi (t + i\beta)[B_\ell]\rangle  =
 \text{Tr} \left[\rho_\varphi  e^{iH_\varphi t }\; B \; e^{-iH_\varphi t }A
\right] =  \langle \varphi \; ; \;  \alpha_\varphi (t)[B_\ell]\;A_\ell\rangle\; , $$
which is the KMS condition.

\subsection{The  antilinear operator $S_\varphi$}\label{sec:antilin-op}
  We now analyze the antilinear operator $S_\varphi : \hs \longrightarrow \hs$, which acts as
(see (\ref{mod-antilin-map}))
\be
  S_\varphi (A_\ell \bPhi ) = A_\ell^* \bPhi \; , \qquad \forall A_\ell \in \mathfrak A_\ell\; .
\label{antilinop}
\en
Taking $A_\ell = A\vee I$,
$$
S_\varphi (A_\ell \bPhi ) = A_\ell^* \bPhi \; , \quad \forall A_\ell \in \mathfrak A_\ell
\quad \Longleftrightarrow \quad
S_\varphi (A\bPhi ) = A^* \bPhi \; , \quad \forall A \in \lh\; . $$
Using (\ref{KMS-vect}) we may write,
$$ S_\varphi (A\bPhi ) = A^* \bPhi \quad \Longrightarrow \quad
\sum_{i=1}^N \alpha_i^{\frac 12}S_\varphi (A\mathbb P_i ) =
\sum_{i=1}^N \alpha_i^{\frac 12} A^* \mathbb P_i \; . $$
Taking  $A = X_{k \ell}$ (see (\ref{basis-vects})) and using $X_{k\ell}\mathbb P_i
= \delta_{\ell i }X_{ki}$, we then get
\be
  \alpha_\ell^{\frac 12}S_\varphi (X_{k\ell} ) = \alpha_k^{\frac 12}S_\varphi (X_{\ell  k} )
  \quad \Longrightarrow \quad  S_\varphi (X_{k\ell} ) =
  \left[\frac { \alpha_k}{ \alpha_\ell}\right]^{\frac 12}X_{\ell k}\; .
\label{antiliop2}
\en
Since any $A \in \lh$ can be written as $A = \sum_{i,j =1}^N a_{ij}X_{ij}$, where
$a_{ij} = \langle \zeta_i \mid A\zeta_j \rangle$, and furthermore, since
$ \bP_{ij} (X_{k\ell}) = X_{ij}\delta_{ik}\delta_{j\ell}$, we obtain using
(\ref{J-op}) and  (\ref{delta-op}),
\be
   S_\varphi = J [\Delta_\varphi]^{\frac 12}\; ,
\label{antilinop3}
\en
which in fact, also gives the polar decomposition of $S_\varphi$.

  Thus, we could have obtained, as described in Section \ref{sec:summ-math},
 the time evolution automorhisms
$\alpha_\varphi (t)$, $ t\in \mathbb R$, by analyzing the  antilinear operator $S_\varphi$,
(since $S_\varphi ^* S_\varphi = \Delta_\varphi$) directly.  Also, from (\ref{hamiltonian1}),
(\ref{delta-op}) and (\ref{timeev-op2}) we see that
the modular operator simply defines the {\em Gibbs state} corresponding to the Hamiltonian
 $\bH_\varphi$.

\subsection{The centralizer}\label{sec:centralizer}
As defined in the Appendix, the centralizer of $\mathfrak A_\ell$, with respect to the
state $\varphi$, is the von Neumann algebra,
\be
  \mathfrak M_\varphi = \{B_\ell \in \mathfrak A_\ell \mid \langle \varphi \; ; \; [B_\ell ,
  A_\ell ] \rangle = 0\; , \;\; \forall A_\ell \in \mathfrak A_\ell \}\; .
\label{centralizer}
\en
Let us determine this von Neumann algebra. Writing $A_\ell = A\vee I\; , \;\;
B_\ell = B \vee I$, the commutator,  $[B_\ell , A_\ell ] = (AB -BA)\vee I$. Hence, by
(\ref{vect-state}),
$$ \langle \varphi \; ; \; [B_\ell ,  A_\ell ] \rangle =
             \text{Tr} \left[ \rho_\varphi (AB - BA)\right]\; .$$
Thus, in order for the above expression to vanish, we must have,
$$ \sum_{i=1}^N \alpha_i \langle \zeta_i \mid AB \zeta_i \rangle =
\sum_{i=1}^N \alpha_i \langle \zeta_i \mid BA \zeta_i \rangle\; ,
\qquad \forall A \in \lh\; .$$
Taking $A = \vert \zeta_k\rangle\langle \zeta_\ell \vert$, this gives,
$$ \alpha_k \langle \zeta_\ell \mid B \zeta_k\rangle =
    \alpha_\ell \langle \zeta_\ell \mid B \zeta_k\rangle\; ,
    \qquad \forall k, \ell = 1,2, \ldots , N\; ,$$
and since in general, $\alpha_k \neq \alpha_\ell$, this implies that
$\langle \zeta_\ell \mid B \zeta_k\rangle = 0$ whenever $k \neq \ell$. Thus, $B$ is
of the general form $B = \sum_{i=1}^N b_i \mathbb P_i\; , \;\; b_i \in \mathbb C$.
In other words, the centralizer
$\mathfrak M_\varphi$ is generated by the projectors $\mathbb P_i^\ell =
\mathbb P_i \vee I\; ,
i=1,2, \ldots , N$,  which are {\em minimal} (i.e., they do not contain projectors onto
smaller subspaces) in $\mathfrak A_\ell$.  Alternatively, we may write,
$\mathfrak M_\varphi = \{H_\varphi^\ell\}^{\prime \prime}$, where $H_\varphi^\ell$ is the
Hamiltonian defined in  (\ref{hamiltonian2}), so that it is an {\em atomic,
commutative} von Neumann algebra.

\section{Application to Landau levels}\label{sec:applications}
  We now show how the above setup, based on $\hs$, can be applied to a specific physical
situation namely, to the case of  an electron subject to a constant magnetic
field, as discussed in \cite{alibag}.

In that case, $\h = L^2 (\mathbb R)$ and
the mapping $\mathcal W : \hs  \longrightarrow L^2 (\mathbb R^2, dx\;dy)$, with
\be
   (\mathcal W X)(x,y) = \frac 1{(2\pi)^{\frac 12}} \text{Tr}[ U(x, y)^* X], \quad
\text{where} \quad U(x,y) = e^{-i(xQ + yP)},
\label{wigmap}
\en
$Q, P$ being the usual position and momentum operators ($[Q,P] = iI$), transfers the whole
modular structure unitarily to the Hilbert space $\htil = L^2 (\mathbb R^2, dx\;dy)$. The
mapping $\mathcal W$ is often referred to as the {\em Wigner transform} in the physical
literature.

To work this out in some detail, we start by  constructing the
Hamiltonian $H_\varphi$ (see (\ref{hamiltonian1})),  using the oscillator
Hamiltonian $H_\text{osc} = \frac 12 (P^2 + Q^2)$ on $\h$.  Let us  choose the
orthonormal basis set of vectors $\zeta_n\; , n =0,1, 2, \ldots \infty$, to be the
eingenvectors of $H_\text{osc}$:
\be
  H_\text{osc} \zeta_n = \left(n + \frac 12 \right) \zeta_n\; .
\label{osc-eigenfunc}
\en
As is well known, the $\zeta_n$ are the Hermite functions,
\be
  \zeta_n (x) = \frac 1{\pi^{\frac 14}}\frac 1{\sqrt{2^n \;n!}} \;e^{-\frac {x^2}2} h_n (x)\; ,
\label{herm-func}
\en
 the $h_n$ being the Hermite polynomials, obtainable as:
\be
  h_n (x) = (-1)^n \; e^{x^2} \partial^n_x \; e^{- x^2}\; .
\label{real-herm-poly}
\en

Consider now the operator $e^{-\beta H_\text{osc}}$, for some fixed $\beta > 0$. We have,
$$
   e^{-\beta H_\text{osc}} = \sum_{n=0}^\infty e^{-(n + \frac 12)\beta}\mathbb P_n
 \quad \text{and} \quad  \text{Tr} \left[ e^{-\beta H_\text{osc}} \right ] =
 \frac {e^{-\frac {\beta}2}}{1 - e^{-\beta}}\; .  $$
 Thus we take,
\be
  \rho_\varphi = \frac {e^{-\beta H_\text{osc}}}{\text{Tr}
  \left[ e^{-\beta H_\varphi} \right ]} =
  (1 - e^{-\beta})\sum_{n=0}^\infty e^{-n\beta}\mathbb P_n\; \quad \text{and}
  \quad \bPhi = \left[1 - e^{-\beta}\right]^{\frac 12} \sum_{n=0}^\infty
  e^{-{\frac n2}\beta}\mathbb P_n\; .
\label{KMS4}
\en
Following  (\ref{vect-state}) and (\ref{hamiltonian1}) we write,
$$
    \rho_\varphi = \sum_{n=0}^\infty \alpha_n \mathbb P_n \; , \qquad
    \alpha_n =  (1 - e^{-\beta})e^{-n\beta}\; , $$
and
\bea
  H_\varphi & =&  - \frac 1\beta \sum_{n=0}^\infty \log\left[(1 - e^{-\beta})e^{-n\beta}\right]
      \mathbb P_n
    =    \sum_{n=0}^\infty \left[ n - \frac { \log(1 - e^{-\beta})}\beta \right]
      \mathbb P_n \nonumber\\
     & = &  H_\text{osc} - \left[\frac 12 +  \frac {\log(1 - e^{-\beta})}\beta \right]  I\; ,
\label{modhamilt}
  \ena
which is the Hamiltonian giving the time evolution  $\alpha_\varphi (t)$,  with respect
to which the above $\rho_\varphi$ defines the KMS state $\varphi$. Since the difference
between  $H_\varphi$ and $H_{\text{osc}}$ is just a constant,
we shall identify these two Hamiltonians in the sequel.

  As stated earlier, the dynamical model that we consider is that of a single electron
of unit charge, placed in the $xy$-plane and subjected to a constant magnetic
field, pointing along the {\em positive $z$-direction\/}. The classical
Hamiltonian of the system, in some convenient units, is
\be
  H_\text{elec} = \frac 12 (\vec p - \vec A )^2 = \frac 12 \left(p_x + \frac y2 \right)^2 +
      \frac 12 \left(p_y - \frac x2 \right)^2\; ,
\label{elec-ham}
\end{equation}
where we have chosen the magnetic vector potential to be $\vec A^\uparrow := \vec A =
\frac 12 (-y, x, 0)$ (so that the magnetic field, $\vec B = \nabla \times \vec A^\uparrow
= (0,0, 1)$).

Next, on  $\htil = L^2 (\mathbb R^2 , dxdy )$, we introduce the quantized
observables, \be
 p_x + \frac y2 \longrightarrow Q_- = -i\frac \partial{\partial x } + \frac y2 \; , \qquad
 p_y - \frac x2 \longrightarrow P_- = -i\frac \partial{\partial y } - \frac x2 \; ,
\label{quant-obs1}
\end{equation}
which satisfy $[Q_-, P_- ] = iI_{\htil}$ and in terms of which the quantum Hamiltonian,
corresponding to $H_\text{elec}$ becomes
\be
  H^\uparrow = \frac 12 \left( P_-^2 + Q_-^2\right)\; .
\label{qu-elec-ham1}
\end{equation}
This is the same as the oscillator Hamiltonian in one dimension, $H_{\text{osc}}$,
given above (and hence the same as $H_\varphi$ in (\ref{modhamilt}), with our convention
of identifying these two). The eigenvalues of this Hamiltonian are then
$E_\ell = (\ell + \frac 12 ), \;
\ell =0,1,2, \ldots \infty$. However, this time each level is infinitely degenerate,
and we will denote the corresponding normalized eigenvectors by
$\Psi_{n \ell}$, with $\ell = 0, 1,2, \ldots , \infty$, indexing  the energy level and
$n = 0,1,2, \ldots , \infty$, the degeneracy at each level. If the magnetic field
were aligned along the {\em negative
$z$-axis} (with $\vec A^\downarrow = \frac 12 (y, -x, 0)$ and
$\vec B = \nabla \times \vec A^\downarrow = (0,0, -1)$), the corresponding quantum
Hamiltonian would have been
\be
  H^\downarrow = \frac 12 \left( P_+^2 + Q_+^2\right)\; .
\label{qu-elec-ham2}
\end{equation}
with
\be
 Q_+ = -i\frac \partial{\partial y } + \frac x2 \; , \qquad
 P_+ = -i\frac \partial{\partial x } - \frac y2 \; ,
\label{quant-obs2}
\end{equation}
and $[Q_+, P_+ ] = iI_{\htil}$. The two sets of operators $\{ Q_\pm , P_\pm\},$
mutually commute:
\be
  [Q_+ , Q_- ] = [P_+ , Q_- ] = [Q_+ , P_- ] = [P_+ , P_- ] = 0 \; .
\label{commutants}
\end{equation}

Thus, $[H^\downarrow , H^\uparrow ] = 0$ and the eigenvectors $\Psi_{n \ell}$
of $H^\uparrow$ can be  chosen so that they are
also the eigenvectors of $H^\downarrow$ in the manner
\be
  H^\downarrow\Psi_{n\ell} =  \left(n + \frac 12 \right)\Psi_{n\ell} \; , \qquad
  H^\uparrow\Psi_{n\ell} =  \left(\ell + \frac 12 \right)\Psi_{n\ell}\; ,
\label{deg-lift}
\end{equation}
so that $H^\downarrow$ lifts the degeneracy of $H^\uparrow$ and vice versa.
In what follows, we  shall assume that this is the case.

Then, it is well known (see, for example, \cite{aag_book}) that the  map $\mathcal W$ in
(\ref{wigmap})
is unitary and straightforward computations (see, for example \cite{alibag}) yield,
\be
 \mathcal W \begin{pmatrix} Q\vee I_\h \\ P\vee I_\h \end{pmatrix}\mathcal W ^{-1} =
 \begin{pmatrix} Q_+ \\ P_+ \end{pmatrix}\; , \qquad
 \mathcal W \begin{pmatrix} I_\h \vee Q \\ I_\h \vee P \end{pmatrix} \mathcal W^{-1} =
 \begin{pmatrix} Q_- \\ P_- \end{pmatrix}\;,
\label{alg-transf1}
\end{equation}
and
\be
\mathcal W \begin{pmatrix} H_\text{osc}\vee I_\h \\ I_\h\vee H_\text{osc}
\end{pmatrix}\mathcal W^{-1} =
 \begin{pmatrix} H^\downarrow \\ H^\uparrow \end{pmatrix}\; , \qquad \mathcal W  X_{n\ell} = \Psi_{n\ell },
\label{alg-transf2}
\end{equation}
where the $X_{n\ell}$ are the basis vectors defined in
(\ref{basis-vects}) and the $\Psi_{n\ell}$ are the normalized
eigenvectors defined in (\ref{deg-lift}). This also means that
these latter vectors form a basis of $\htil = L^2(\mathbb R^2 , dxdy)$. Finally, note that the
two sets of operators, $\{Q_+ , P_+  \}$ and $\{Q_- , P_- \}$, generate (see Appendix)
the two von Neumann algebras
$\mathfrak A_+$ and $\mathfrak A_-$, respectively, with $\mathcal W \mathfrak A_\ell
\mathcal W^{-1} =
\mathfrak A_+$ and $\mathcal W \mathfrak A_{\text{r}}\mathcal W^{-1}  = \mathfrak A_-$.
Thus physically, the two commuting algebras correspond to the two directions of the
magnetic field. The KMS state $\bPsi = \mathcal W \bPhi$, with $\bPhi$ given by
(\ref{KMS4}) is just the {\em Gibbs equilibrium state} for this physical system.

\section{A second representation}\label{sec:second-rep}
  It is interesting to pursue this example a bit further by transforming to
complex coordinates, which will essentially reduce the action of the
operator $J$ to one of complex conjugation.  The
possibility of having this other representation is a reflection of the fact that there
is more than
one possible way to represent  the two commuting von Neumann algebras
$\mathfrak A_\pm$.
As before, let us consider the electron in a uniform magnetic field
oriented in the positive $z$-direction, with vector potential
$\vec A^\uparrow =\frac{1}{2}(-y,x,0)$ and magnetic field $\vec B = \nabla \times \vec A^\uparrow
= (0,0, 1)$). The classical Hamiltonian is now given by
$H^\uparrow=\frac{1}{2}\left(\vec p- \vec A^\uparrow\right)^2$.
There are several possible ways to write this Hamiltonian, which are more convenient
than using the coordinates $x, y$ and $z$. One such representation was used in
Section \ref{sec:applications} and we indicate below a second possibility. Note that the
quantized Hamiltonian may be split into a free part $H_0$ and an interaction
or angular momentum part,
$H_{\text{int}}^\uparrow$:
\be
\left\{
\begin{array}{ll}
H^\uparrow=H_0+H_{\text{int}}^\uparrow,\\[2pt]
H_0=H_{0,x}+H_{0,y}=\frac{1}{2}\left( \widehat{p}_x^2+ \dfrac {\widehat{x}^2}4\right)+
\frac{1}{2}\left(\widehat{p}_y^2+ \dfrac {\widehat{y}^2}4 \right),\\[2pt]
H_{\text{int}}^\uparrow=- \dfrac 12 (\widehat{x}\widehat{p}_y- \widehat{y}\widehat{p}_x)
    =-\widehat{l}_z\; .
\end{array}
\right.\label{11}\en
with the usual definitions of $\widehat{x} , \widehat{p}_x\; ,$ etc. Of
course, $[\widehat{x}, \widehat{p}_x]=[\widehat{y} ,\widehat{p}_y]=iI_{\htil}$, while all
the other commutators are zero. Introducing the corresponding annihilation operators,
\be
a_x= \frac 1{\sqrt{2}} [\widehat{x} + i\widehat{p}_x ]\;,\qquad
a_y= \frac 1{\sqrt{2}} [\widehat{y} + i\widehat{p}_y ]\; ,
\label{12}
\en
and their adjoints,
\be
a_x^* = \frac 1{\sqrt{2}} [\widehat{x} -  i\widehat{p}_x]\; ,\qquad
a_y^* = \frac 1{\sqrt{2}} [\widehat{y} - i\widehat{p}_y] \; ,
\label{122}
\en
which satisfy the canonical commutation rules
$[a_x,a_x^*]=[a_y,a_y^*]=I_{\htil}$, while all the other commutators are zero,
the hamiltonian $H^\uparrow$ can be written as
$H^\uparrow=H_0+H_{\text{int}}^\uparrow$, with $H_0= \left(a_x^*a_x+a_y^*a_y+ I_{\htil}\right)$,
$H_{\text{int}}^\uparrow=-i(a_x a_y^*-a_ ya_x^*)$. $H^\uparrow$ does not appear to be
diagonal even in this form, so that another {\em change of variables} is required.

  Using the operators $Q_\pm , P_\pm$, given in (\ref{quant-obs1}) and (\ref{quant-obs2}),
let us define
\bea
  A_+  & =  & \frac 1{\sqrt{2}} (Q_+ + iP_+)  = \frac 34 (a_x - ia_y) - \frac 14 (a_x^* + ia_y^* )\; ,  \nonumber\\
  A_+^*  & = & \frac 1{\sqrt{2}} (Q_+ - iP_+)  = \frac 34 (a_x^* + ia_y^*) - \frac 14 (a_x - ia_y )\; ,
  \nonumber\\
 A_- & = &  \frac 1{\sqrt{2}} (iQ_- - P_-)  = \frac 34 (a_x + ia_y) - \frac 14 (a_x^* + ia_y^* )\; , \nonumber\\
   A_-^* &  = & \frac 1{\sqrt{2}} (-iQ_- - P_-)  = \frac 34 (a_x^* - ia_y^*) - \frac 14 (a_x - ia_y ) \; .
\label{new-ccr}
\ena

These satisfy the commutation relations,
\be
 [A_\pm , A^*_\pm ] = 1\; ,
\label{new-ccr2}
\en
with all other commutators being zero. In terms of these, we may write the
two Hamiltonians as (see (\ref{qu-elec-ham1}) and (\ref{qu-elec-ham2}),
\be
  H^\uparrow =  N_- + \frac 12 I_{\htil} \; , \quad
  H^\downarrow = N_+ + \frac 12 I_{\htil}\;, \quad \text{with} \quad N_\pm =
  A^*_\pm A_\pm\; .
\label{qu-elec-ham6}
\en
Furthermore,
\be
  H_0 = \frac 12 (N_+ + N_- +1) \quad \text{and} \quad H_{\text{int}}^\uparrow
    =  -\frac 12 (N_+ - N_-)\; , \quad H_{\text{int}}^\downarrow
    =  \frac 12 (N_+ - N_-)\; .
\label{qu-elec-ham7}
\en

   The eigenstates of $H^\uparrow$
are now easily written down. Let $\Psi_{00}$ be such that $A_-\Psi_{00}=A_+\Psi_{00}=0$.
Then we define
\be
\Psi_{n \ell}:=\frac{1}{\sqrt{n!\ell!}}\left(A_+^*\right)^n\left(A_-^*\right)^\ell\Psi_{0 0},
\label{15}
\en
where $n,\ell =0,1,2,\ldots$. All the relevant operators are now diagonal in this basis:
$N_+\Psi_{n \ell }=n\Psi_{n \ell }\;,\;\; N_-\Psi_{n \ell }=\ell \Psi_{n \ell }\;,\;\;
H_0\Psi_{n \ell }= \frac 12
(n+\ell+1)\Psi_{n \ell}$ and $H_{\text{int}}^\uparrow\Psi_{n \ell }=
\frac 12 (n-\ell )\Psi_{n \ell}$. Hence
\be
H^\uparrow\Psi_{n \ell }= \left(\ell +\frac{1}{2}\right)\Psi_{n \ell}
\label{16}
\en
This means that each level $\ell$ is infinitely degenerate, with $n$ being the  degeneracy
index. Again, this degeneracy can be lifted in a
physically interesting way namely, by considering the {\em reflected} magnetic field with
vector potential $\vec A^\downarrow=\frac{1}{2}(y,-x,0)$ as in Section \ref{sec:applications},
with the magnetic field  directed along the negative $z$-direction.  The same electron
considered above is now described by the other  Hamiltonian,
$H^\downarrow$, which can be written as
\be
\left\{
\begin{array}{ll}
H^\downarrow=\frac{1}{2}\left(\vec p - \vec A^\downarrow\right)^2=
H_0+H_{\text{int}}^\downarrow,\\
H_{\text{int}}^\downarrow=-H_{\text{int}}^\uparrow
\end{array}
\right.\label{17}
\en
Thus, since  $H^\downarrow$ can also be written as  in (\ref{qu-elec-ham6}), its
eigenstates are again  the same $\Psi_{n \ell }$ given in (\ref{15}). (Recall that
 $[H^\uparrow,H^\downarrow]=0$, so that they can be simultaneously
diagonalized.) This also means that, as in (\ref{alg-transf2}), $\mathcal W X_{n\ell}
=\Psi_{n\ell}$ and the closure of the linear span of the $\Psi_{n \ell}$'s is the Hilbert space
$\htil = L^2 (\mathbb R^2 , dx\;dy )$.

However, this is not the end of the story. Indeed, introducing the complex variable
$z=\dfrac 1{\sqrt{2}}(x - iy)$ and its associated derivative
$\partial_z =
\dfrac{1}{\sqrt{2}}(\partial_x + i\partial_y)$,
the operators $A_+$ and $A_-$ can be written as
\be
A_- = \frac 12 \overline{z}+ \partial_z \; ,
\qquad A_+= \frac 12 z + \partial_{\overline z}\;,
\label{complex-ops1}
\en
and their adjoints as,
\be
A_-^* = \frac 12 z - \partial_{\overline z} \; ,
\qquad A_+^* = \frac 12 {\overline z}  -  \partial_z\;,
\label{complex-ops2}
\en

In this $(\overline{z}, z)$-representation the ground state $\Psi_{00}(\overline{z}, z)$
is the solution of the equations $A_+\Psi_{00}(\overline{z}, z)=
 A_-\Psi_{00}(\overline{z}, z)
=0$, so that,  $\Psi_{00}(\overline{z}, z )=\sqrt{\dfrac{1}{2\pi}}\;e^{-\frac{1}{2}|z|^2}$.
We conclude from (\ref{alg-transf2}) and (\ref{15}) that
\be
\Psi_{n \ell }(\overline{z}, z) = (\mathcal W X_{n\ell})(\overline{z}, z)
  =  \frac{1}{\sqrt{n!\ell!}} \left(\frac 12
\overline{z} - \partial_z\right)^n \left(\frac 12 z -
\partial_{\overline z}\right)^\ell \Psi_{00}(z,\overline{z})\; .
\label{18}
\en

When using this representation, we shall write our Hilbert space as $\htil =
L^2(\mathbb C , \frac {d\overline{z}\wedge dz}{i} )$ and then it is
useful to make the further unitary transformation
\be
\mathcal V : L^2(\mathbb C , \frac {d\overline{z}\wedge dz}{i} )
   \longrightarrow L^2 (\mathbb C , d\nu (\overline{z} , z ))
\quad \text{where} \quad d\nu (\overline{z} , z ) = \frac {e^{-\vert z\vert^2}}{2\pi} \;
  \dfrac {d\overline{z}\wedge dz}{i} \; ,
\label{hol-transf1}
\en
to the more convenient Hilbert space $L^2 (\mathbb C , d\nu (\overline{z} , z ))$, using
the mapping
\be
  (\mathcal V \Psi )(\overline{z} , z ) = \sqrt{2\pi}\; e^{\frac {\vert z\vert^2}2}\;
  \Psi(\overline{z} , z )\; ,
  \label{hol_trasf2}
  \en
and to rewrite all the operators in question on this new space. Note that this space
contains the two subspaces $\h_{\text{hol}}$ and $\h_{\text{a-hol}}$, of holomorphic
and antiholomorphic functions, respectively. Both these subspaces contain the constant
unit vector, $H_{00} (\overline{z}, z ) = 1\; , \; \forall (\overline{z} , z)$. Apart
from this one vector, all other vectors in the complementary subspaces of
$\h_{\text{hol}}$ and $\h_{\text{a-hol}}$ are mutually orthogonal.
Since
$$ \frac {\partial}{\partial z}\left[\Psi (\overline{z}, z )\;e^{\frac {\vert z\vert^2}2}\right]
  = \left[\frac {\partial}{\partial z}\Psi (\overline{z}, z )
    + \overline{z} \Psi (\overline{z}, z )\right]\;e^{\frac {\vert z\vert^2}2}\; , $$
we immediately find that
\bea
\mathcal A_- := \mathcal V A_-\mathcal V^{-1} =  \partial_z \; ,
&\qquad& \mathcal A_+ := \mathcal V A_+\mathcal V^{-1} =\partial_{\overline z}\;,
\nonumber\\
\mathcal A_-^* := \mathcal V A_-^* \mathcal V^{-1} = z- \partial_{\overline z} \; ,
&\qquad& \mathcal A_+^* := \mathcal V A_+^*\mathcal V^{-1} = \overline{z} - \partial_z\;,
\label{complex-ops6}
\ena
Furthermore, in this representation we have the number operators,
\bea
\mathcal N_+  & := & \mathcal V N_+\mathcal V^{-1} = \mathcal A_+^* \mathcal A_+ =
- \partial_z \partial_{\overline z}
    + \overline{z}\partial_{\overline z}\; \nonumber\\
\mathcal N_-  & := & \mathcal V N_-\mathcal V^{-1} = \mathcal A_-^* \mathcal A_-
 =  - \partial_z \partial_{\overline z}
   + z \partial_z \;.
\label{hol-numops1}
\ena

   Writing $H_{n\ell }= \mathcal V \Psi_{n \ell }$, for the transformed basis vectors
(\ref{18}), we have,
\bea
H_{n\ell }(\overline{z}, z)  & =  &   \frac{1}{\sqrt{n!\ell!}} \left(
\overline{z} - \partial_z\right)^n \left( z -
\partial_{\overline z}\right)^\ell H_{00}(\overline{z}, z)
  =  \frac{1}{\sqrt{n!\ell!}}\left( \mathcal A_+^* \right)^n
 \left( \mathcal A_-^* \right)^\ell H_{0 0}(\overline{z}, z)\; ,\nonumber\\
 & = &  \frac{1}{\sqrt{n!\ell!}} \left(
\overline{z} - \partial_z\right)^n (z^\ell ) =  \frac{1}{\sqrt{n!\ell!}}  \left( z -
\partial_{\overline z}\right)^\ell(\overline{z})^n\; .
\label{hol-basis}
\ena
Also,
\be
  H_{n0}(\overline{z}, z) = \frac {\overline{z}^n}{\sqrt{n!}} \quad \text{and}
  \quad H_{0\ell}(\overline{z}, z) = \frac {z^\ell}{\sqrt{\ell!}}\; ,
\label{hol-basis7}
\en
so that $\h_{\text{a-hol}}$ is spanned by the vectors $H_{n0}\; , n =0,1,2, \ldots , $
and the space $\h_{\text{hol}}$ by the vectors $H_{0\ell}\; , \;\ell =0,1,2, \ldots \; $.

The vectors $H_{n\ell}$ are joint eigenstates of the number operators:
\be
\mathcal N_+ H_{n\ell } = \ell H_{n\ell }\; , \qquad \mathcal N_- H_{n\ell}
= n H_{n\ell}\;.
\label{hol-numops2}
\en
Moreover, writing
\bea
 \mathcal H^\uparrow = \mathcal V H^\uparrow \mathcal V^{-1} \; , &\qquad&
   \mathcal H^\downarrow = \mathcal V H^\downarrow \mathcal V^{-1} \;,\nonumber\\
  \mathcal H_0 = \mathcal V H_0 \mathcal V^{-1} \; , &\qquad&
   \mathcal H^{\uparrow,\;\downarrow}_{\text{int}} =
   \mathcal V H^{\uparrow,\;\downarrow}_{\text{int}} \mathcal V^{-1} \;,
\label{hol-hams}
\ena
for the two Hamiltonians, we clearly have (see (\ref{16})),
\be
\mathcal H^\uparrow H_{n\ell} = \left(\ell + \frac 12\right)H_{n\ell}\;, \qquad
\mathcal H^\downarrow H_{n\ell} = \left(n + \frac 12\right)H_{n\ell}\; .
\label{hol-hams2}
\en

   The functions  $h_{n,k} (\overline{z} , z) = \sqrt{n! k!}\; H_{nk}(\overline{z} , z)$
are just the {\em complex Hermite polynomials} \cite{intint,ghan}, also obtainable as:
\be
 h_{n,k} (\overline{z} , z)  = (-1)^{n+k} \;e^{\vert z\vert^2}\partial^n_z \partial^k_{\overline z}
   \; e^{-\vert z\vert^2}\; .
\label{comp-herm-poly}
\en
Explicitly, the $h_{n,k}$ are given by
\be
  h_{n,k} ( \overline{z} , z ) = n!\;k! \sum_{j= 0}^{n\curlyvee k}
    \frac {(\overline{z})^{n-j}}{(n-j)!}\; \frac {z^{k-j}}{(k-j)!}\; ,
\label{comp-herm-poly9}
\en
where $n\curlyvee k$ denotes the smaller of the two numbers $n$ and $k$.
In particular,
\be
 h_{0,k} (\overline{z} , z) = z^k \quad \text{and} \quad
 h_{n,0} (\overline{z} , z) = \overline{z}^n\; .
\label{comp-herm-poly2}
\en
One also has the useful series expansion,
\be
\sum_{n=0}^\infty \sum_{k=0}^\infty \frac {v^n\overline{u}^k }{n!k!}h_{n,k}(\overline{z} , z )
=\sum_{n=0}^\infty \sum_{k=0}^\infty
 \frac {v^n\overline{u}^k }{\sqrt{n!k!}}H_{nk}(\overline{z} , z )
  = e^{\overline{u} z + v \overline{z} - \overline{u} v}\; .
\label{comp-herm-poly12}
\en

 Furthermore,
\be
h_{n,k} (\overline{z} , z) = h_{k,n} (z, \overline{z}) \quad \text{and} \quad
h_{n,k} (\overline{z} , z) = ((\mathcal A^*_+ )^n h_{0, k}) (\overline{z} , z )\; .
\label{comp-herm-poly3}
\en
They also satisfy the recursion relations,
\bea
h_{n+1,k} (\overline{z} , z) & = & \overline{z}\; h_{n,k} (\overline{z} , z) -
                                 k \;h_{n,k-1} (\overline{z} , z)\nonumber\\
h_{n,k+1} (\overline{z} , z) & = & z \;h_{n,k} (\overline{z} , z) -
                                n\;h_{n-1,k} (\overline{z} , z)\; ,
\label{comp-herm-poly4}
\ena
from which we further obtain
\bea
  \overline{z}\; h_{n, n+1} (\overline{z} , z) & = &
        z\; h_{n+1, n} (\overline{z} , z)\nonumber\\
(k-m)\;h_{m,k} (\overline{z} , z) & = &  \overline{z}\; h_{m, k+1} (\overline{z} , z)
    -  z\; h_{m+1 , k} (\overline{z} , z)\; .
\label{comp-herm-poly5}
\ena

If we formally take z to be real in (\ref{comp-herm-poly}), the complex Hermite
polynomials $h_{n,k} (\overline{z} , z) $
become the well-known real Hermite polynomials $h_{n+k} (x)$ (see (\ref{real-herm-poly})) :
$$
  h_n (x) = (-1)^n \; e^{x^2} \partial^n_x \; e^{- x^2}\; ,
$$
which satisfy the recursion relations,
\be
   x h_n (x) = n h_{n-1}(x) + \frac 12 h_{n+1}(x)\; .
\label{real-herm-poly2}
\en

\section{Some physical considerations}\label{sec-phys-disc}
    The composite map
\be
  \mathcal U := \mathcal V \circ \mathcal W : \mathcal B_2 (\h ) \longrightarrow
  L^2 (\mathbb C , d\nu (\overline{z} , z ))\; , \quad \text{with}
  \quad \mathcal U X_{n\ell} = H_{n\ell}\; ,
\label{comp-map}
\en
transforms the modular conjugation map $J$ in (\ref{J-op}) to $\mathcal J =
\mathcal U J \mathcal U^{-1}$ which  basically acts by complex conjugation
(see (\ref{complex-ops6}) and (\ref{comp-herm-poly3})):
\be
  (\mathcal J H_{n\ell}) (\overline{z} , z ) = H_{\ell n} (\overline{z} , z ) =
 H_{n  \ell} (z , \overline{z}) \quad \text{and} \quad
  \mathcal J \mathcal A_\pm \mathcal J^{-1} = \mathcal A_\mp =
  \overline{\mathcal A_\pm }\; ,
\label{mod-j-op}
\en
etc. (The overline indicates complex conjugation of the variables appearing in the
definitions of the operators). Similarly, the two mutually commuting algebras,
$\mathfrak U_+ = \mathcal U \;\mathfrak A_+ \mathcal U^{-1}$ and
$\mathfrak U_-  = \mathcal U\; \mathfrak A_- \mathcal U^{-1}$, generated by the
two sets of operators $\{\mathcal A_+ , \mathcal A^*_+\}$ and
$\{\mathcal A_- , \mathcal A^*_-\}$, respectively, are
transformed into each other  by $\mathcal J$, both algebras being factors. Additionally,
\be
\mathcal J \mathcal H_0 \mathcal J = \mathcal H_0, \qquad
\mathcal J\mathcal H_{\text{int}}^\uparrow \mathcal J =
\mathcal H_{\text{int}}^\downarrow=-\mathcal H_{\text{int}}^\uparrow\quad \Rightarrow
\quad \mathcal J \mathcal H^\uparrow \mathcal J =\mathcal H^\downarrow\; .
\label{add34}
\en
The KMS state on the algebra $\mathfrak U_+$, which
is a vector state,
is given by the vector (see (\ref{KMS-vect}) and (\ref{KMS4})),
\be
  {\bcalX} = \mathcal U \bPhi = (1-e^{-\beta})^{\frac 12}
    \sum_{n=0}^\infty e^{-\frac {n\beta}2}H_{nn} \in
    L^2 (\mathbb C , d\nu (\overline{z}, z))\; , \qquad \mathcal J \bcalX = \bcalX\; .
\label{KMS13}
\en
The Hamiltonian
\be
   \bcalH_\varphi = \mathcal H^\uparrow - \mathcal H^\downarrow =
   2\mathcal H_{\text{int}}^\uparrow
        = - 2(\mathcal N_+ - \mathcal N_- )\; ,
\label{mod-ham2}
\en
then gives the modular operator,
\be
 \Delta_\varphi = \exp [-\beta\bcalH_\varphi] = \sum_{n,\ell = 0}^\infty e^{-\beta (n- \ell)}\vert H_{n\ell}\rangle
      \langle H_{n\ell}\vert\; ,
\label{mod-op16}
\en
and the one-parameter automorphism group,
\be
    \Delta_\varphi^{-\frac {it}\beta} =   \exp [i\bcalH_\varphi t]
    = \exp [2i\mathcal H_{\text{int}}^\uparrow t ]\; , \;\; t \in \mathbb R\; ,
\qquad  \Delta_\varphi^{-\frac {it}\beta} \bcalX = \bcalX\; ,
\label{mod-automprph}
\en
which stabilizes $\bcalX$. In other words, the modular automorphism is basically the
time evolution generated by the interaction Hamiltonian. One also verifies that
\be
     \Delta_\varphi^{-\frac {it}\beta}\mathfrak U_+  \Delta_\varphi^{\frac {it}\beta}
         = \mathfrak U_+\; .
\label{mod-automprph2}
\en

The conclusion is therefore the following:
{\em the map $\mathcal J$ in (\ref{mod-j-op})  is, at the same time\/},
\begin{itemize}
\item[$\bullet$] the modular map of the Tomita-Takesaki theory;

\item[$\bullet$] the complex conjugation map;

\item[$\bullet$]  the map which reverses the uniform magnetic field,
  from $\vec B$ to $-\vec B$,
thus transforming  $\mathcal H^\uparrow$ to $\mathcal H^\downarrow$, while
leaving $\mathcal H_0$ unaffected;

\item[$\bullet$] the operator  interchanging the two mutually commuting von Neumann
algebras  $\mathfrak U_\pm$, these latter defining, therefore,  the experimental
setups corresponding to the two directions of the magnetic field;

\item[$\bullet$] an interwining operator in the sense of \cite{intop}, see below.
\end{itemize}

Let us consider this last claim in a bit more detail, following \cite{bagint2009,intop} and
references therein. The main result on this topic is the following: suppose
we have two Hamiltonians, $H_1$ and $H_2$, which are related by an intertwining
operator $X$ in the following way: $XH_1=H_2X$. Then, the knowledge of the
eigensystem of, say, $H_1$, essentially fixes the eigensystem of $H_2$. Indeed we
have  \cite{intop} that, if $\phi^{(1)}_n$ is an eigenstate of $H_1$ with eigenvalue
$E_n$, $H_1\phi^{(1)}_n=E_n\phi^{(1)}_n$, then $X\phi^{(1)}_n$ is either
zero or is an eigenstate of $H_2$ with the same eigenvalue:
$H_2(X\phi^{(1)}_n)=E_n(X\phi^{(1)}_n)$. This is just a consequence of the
intertwining relation. In \cite{bagint2009} this approach has been generalized
proposing a procedure to build up $H_2$ from $H_1$ and from a certain operator
which plays the role of the $X$ above.

 Writing $\mathcal J \mathcal H^\uparrow\mathcal J  =\mathcal H^\downarrow$
 as $\mathcal J \mathcal H^\uparrow
 =\mathcal H^\downarrow \mathcal J$, we see that $\mathcal J$ is an intertwining
 operator between
 $\mathcal H^\downarrow$ and $\mathcal H^\uparrow$. Hence, if $\Phi$ is an
 eigenstate of $\mathcal H^\uparrow$ with eigenvalue $E$, then $\mathcal J\Phi$ is either
 zero or is an eigenvector of $\mathcal H^\downarrow$ with the same eigenvalue.
 This is exactly what is explicitly expressed by equations  (\ref{hol-hams2})
 and (\ref{mod-j-op}) above.

\section{Bi-coherent states and conjugate coherent states}\label{sec:diag-CS}

  There are several natural families of coherent states associated with the Hamiltonians
$\mathcal H^\uparrow , \mathcal H^\downarrow$ and
$\bcalH_\varphi = \mathcal H^\uparrow  -
\mathcal H^\downarrow$. These were constructed in a different context in \cite{alibag}. Here we
look at them again using the complex  Hermite polynomials and the modular conjugation. Using
the expansion in (\ref{comp-herm-poly12}), we define the (non-normalized)
{\em bi-coherent states\/},
\be
  \eta_{\overline{u}, \;v}^{\text{bcs}} =\sum_{n=0}^\infty \sum_{k=0}^\infty
 \frac {v^n\overline{u}^k }{\sqrt{n!k!}}H_{n k} = e^{\frac {\vert u\vert^2 + \vert v\vert^2}2}
  \; e^{\overline{u}\mathcal A_-^* - u\mathcal A_-} \;
   e^{v \mathcal A_+^* - \overline{v} \mathcal A_+}H_{00}\; , \quad u, v \in \mathbb C \; ,
 \label{BCS}
\en
 where use has been made of the following equalities
 $$
    e^{\frac {\vert \alpha\vert^2}2}\; e^{ \alpha\mathcal A^*_+ -\overline{\alpha}\mathcal A_+}
     = e^{ \alpha\mathcal A^*_+}  e^{- \overline{ \alpha}\mathcal A_+}  \quad \text{and} \quad
     e^{\frac {\vert \alpha\vert^2}2}\; e^{ \overline{\alpha} \mathcal A^*_-  - \alpha \mathcal A_-}
     = e^{ \overline{\alpha} \mathcal A^*_-}  e^{ \alpha\mathcal A_-} \; , $$
$\alpha\in\mathbb{C}$. Using the properties of $\mathcal J$ we get
\be
  \mathcal J  \eta_{\overline{u}, \;v}^{\text{bcs}} =
   \eta_{\overline{v},\; u}^{\text{bcs}}\; .
\label{BCS2}
\en
Since the $H_{n k}$ are the eigenfunctions of  $\bcalH_\varphi = \mathcal H^\uparrow  -
\mathcal H^\downarrow = 2\mathcal H^\uparrow_{\text{int}}$, these are coherent states related to the
interaction Hamiltonian, or alternatively,  to the modular automorphism group
(\ref{mod-automprph}). They satisfy the resolution of the identity,
\be
  \int_{\mathbb C}\!\int_{\mathbb C} \vert \eta_{\overline{u},\; v}^{\text{bcs}} \rangle
  \langle \eta_{\overline{u}, \; v}^{\text{bcs}} \vert\; d\nu (\overline{u}, u )\;
   d\nu (\overline{v}, v ) = I_{L^2 (\mathbb C , d\nu (\overline{z} , z ))}\; ,
\label{bcs-resolid}
\en
where $d\nu (\overline{z}, z )$ is the measure introduced in (\ref{hol-transf1}).

   We denote, as before,  the subspace of  $L^2 (\mathbb C,
  d\nu (\overline{z}, z ))$,
consisting of holomorhic functions in the variable $z$,  by $\h_{\text{hol}}$
and  the subspace
of antiholomorphic functions (i.e., holomorphic in $\overline{z}$) by $\h_{\text{a-hol}}$.
As we have seen before, the subspace $\h_{\text{hol}}$
is spanned by the vectors, $\{ H_{00}, H_{01},  H_{02} , $ $\ldots , H_{0n}, \ldots \}$ and
$\h_{\text{a-hol}}$ is spanned by  $\{ H_{00}, H_{10},  H_{20} , \ldots ,
H_{n0}, \ldots \}$.  Let
$\mathbb P_{\text{hol}}$ and $\mathbb P_{\text{a-hol}}$ be the corresponding projection
operators, so that $\mathbb P_{\text{hol}}\cap \mathbb P_{\text{a-hol}} =
\vert H_{00}\rangle\langle H_{00}\vert$.
We now define the (non-normalized) coherent states,
\bea
   \eta_z & : =  &  \eta_{0,\; z}^{\text{bcs}} =
   \sum_{n=0}^\infty \frac { z^n}{\sqrt{n!}}H_{n0} \in \h_{\text{a-hol}}, \quad
         \forall z \in \mathbb C \qquad \Longrightarrow \qquad  \eta_z (\overline{w}) =
         e^{\overline{w}z} \; ,
          \nonumber\\
   \breve{\eta}_{\overline{z}} & : = &  \eta_{\overline{z}, \;0}^{\text{bcs}} =
    \sum_{n=0}^\infty \frac { \overline{z}^n}{\sqrt{n!}}
       H_{0 n}   \in \h_{\text{hol}}, \quad
         \forall \overline{z} \in \mathbb C   \qquad \Longrightarrow \qquad
         \breve{\eta}_{\overline{z}} (w) =
         e^{w\overline{z}} \; .
\label{CS1}
\ena
 Note that this definition is consistent with the fact that the function
$K(\overline{w}, z) = e^{\overline{w}z}$, which
is a reproducing kernel, can be written as (see (\ref{comp-herm-poly2}) and
(\ref{comp-herm-poly12})):
\bea
 e^{\overline{w}z} = \sum_{n=0}^\infty \frac {(\overline{w} z)^n}{n!}&  = &
       \sum_{n=0}^\infty \frac { z^n}{\sqrt{n!}}H_{n 0}(\overline{w}, w ) \nonumber\\
       & = & \sum_{n=0}^\infty \frac { \overline{w}^n}{\sqrt{n!}}H_{0 n}(\overline{z}, z )
\label{kernel1}
\ena

  In view of  (\ref{BCS2}) we also have,
\be
  \mathcal J \eta_z =  \breve{\eta}_{\overline{z}}\; ,
\label{CS2}
\en
and furthermore,
\be
  \eta_z = e^{\frac {\vert z\vert^2}2}\; e^{ z\mathcal A^*_+ -\overline{z}\mathcal A_+}
    H_{0,0}\; ,
  \qquad
 \breve{\eta}_{\overline{z}}  = e^{\frac {\vert z\vert^2}2}\;
    e^{\overline{z}\mathcal A^*_- -  z\mathcal A_- } H_{0,0} \; .
 \label{CS3}
 \en

Because of (\ref{CS2}), we shall call the pair $\{ \eta_z  , \;  \breve{\eta}_{\overline{z}}\}$
{\em conjugate coherent states\/}.

  The following {\em resolutions of the identity}, on $\h_{\text{hol}}$ and
$\h_{\text{a-hol}}$ are then readily established:
\bea
\int_{\mathbb C} \vert\eta_z \rangle\langle \eta_z \vert\; d\nu (\overline{z}, z)
      & = & \mathbb P_{\text{a-hol}}\; , \nonumber\\
\int_{\mathbb C} \vert\breve{\eta}_{\overline{z}} \rangle\langle
      \breve{\eta}_{\overline{z}} \vert\; d\nu (\overline{z}, z)
      & = & \mathbb P_{\text{hol}} \; .
\label{2-resolid}
\ena
Using this and (\ref{CS2}), we also obtain,
\bea
  \mathcal J \;\!\mathbb P_{\text{a-hol}} & = & \int_{\mathbb C} \vert\breve{\eta}_{\overline{z}}
    \rangle\langle \eta_z  \vert\; d\nu (\overline{z}, z)\; ,\nonumber \\
 \mathcal J \;\!\mathbb P_{\text{hol}} & = & \int_{\mathbb C} \vert\eta_z
    \rangle\langle \breve{\eta}_{\overline{z}}\vert \; d\nu (\overline{z}, z)\; .
 \label{cross-resolid}
 \ena
 The first operator is a partial isometry between $\h_{\text{a-hol}}$ and $\h_{\text{hol}}$,
while the second is the reverse isometry.
 Note, also that
 \be
   \mathcal J \;\!\mathbb P_{\text{hol}}\mathcal J = \mathbb P_{\text{a-hol}}\; .
\label{cross-proj}
\en
We also notice that, since (\ref{complex-ops6}) and (\ref{hol-basis}) imply that
$\mathcal A_+H_{n0}=\sqrt{n}\,H_{n-1\; 0}$, we get
$$
\mathcal A_+\eta_z=z\eta_z,\qquad \mathcal A_-\breve{\eta}_{\overline{z}}=
\overline{z}\breve{\eta}_{\overline{z}},
$$
so that, putting
$$\mathbf A=\left(
                      \begin{array}{cc}
                       \mathcal A_+ & 0 \\
                        0 & \mathcal A_- \\
                      \end{array}
                    \right)\; ,  \qquad  \mathcal Z=\left(
                      \begin{array}{cc}
                        z & 0 \\
                        0 & \overline{z} \\
                      \end{array}
                    \right)  \quad \text{and} \quad \bfeta_\mathcal Z=\left(
                      \begin{array}{c}
                        \eta_z  \\
                        \breve{\eta}_{\overline{z}} \\
                      \end{array}
                    \right)\; , $$
 we have $\mathbf A\,\bfeta_{\mathcal Z} =\mathcal Z \,\bfeta_\mathcal Z$. Hence these
 {\em vector coherent states} are eigenvectors of the matrix annihilation operator
 $\mathbf A$ (see also \cite{aeg-vcs}).

\section{Conclusions}\label{sec:concl}
In the problem of the electron studied here, the Hamiltonian governing the motion
had a pure point spectrum, with each level being infinitely degenerate.
Moreover, the energy levels were equally spaced. It seems possible to consider
more general Hamiltonians, again with pure point spectra and infinite
degeneracies, but not equally spaced, and go through a similar analysis. This would lead
to other Hilbert spaces of holomorphic and anti-holomorphic functions and to mutually commuting
von Neumann algebras generated by more general shift operators. Hence a modular structure
similar to that considered in this paper can be recovered.
There are also indications of an
interesting connection between the problem studied here  and the recently studied
non-commutative quantum mechanics, as described, for example,  in  \cite{schgoharo}.
Both these aspects will be considered in a paper which is now in preparation.

\section*{Appendix: basics of von-Neumann algebras}\label{sec-appendix}

In this Appendix we collect together some basic definitions and facts about von Neumann
algebras. Details may, for example, be found in \cite{takesaki2}.
Let $\h$ be a Hilbert space over $\mathbb C$. For
our purposes it will be enough to assume that $\h$ is separable, of dimension $N$, which
could be finite or infinite. We denote by $\mathcal L (\h)$ the set of all bounded
operators on  $\h$. Let $\mathfrak A \subseteq \mathcal L (\h)$ and $\mathfrak A^\prime$
its commutant (i.e., the set of all elements of  $ \mathcal L (\h)$ which commute with
every element of $\mathfrak A$). Let  $\mathfrak A$ be closed under linear combinations,
(operator) multiplication and conjugation (i.e., if $A\in \mathfrak A$ then its
adjoint $A^* \in \mathfrak A$). If in addition $\mathfrak A = \mathfrak A^{\prime\prime}$,
then $\mathfrak A$ is called a {\em von Neumann algebra\/}. It can then be proved that
$\mathfrak A$ is a weakly closed set. A von Neumann algebra always contains the
identity operator $I_\h$ on $\h$. The von Neumann algebra $\mathfrak A$ is called a
{\em factor} if $\mathfrak A \cap \mathfrak A^\prime = \mathbb C I_\h$.

Let   $\varphi : \mathfrak A \longrightarrow
\mathbb C$, be a bounded linear functional on $\mathfrak A$, which we denote
by $\langle\varphi\; ;\; A\rangle\; , \; A\in \mathfrak A$. We call $\varphi$ a {\em state} on
this algebra if it also satisfies the two conditions: $(a)$
$\langle \varphi \; ; \; A^*A\rangle \geq 0\; , \; \forall A \in \mathfrak A$
and  $(b)$ $\langle \varphi \; ; \; I_\h \rangle =1$.
The state $\varphi$ is said to be {\em faithful} if  $\langle \varphi \; ; \; A^*A\rangle > 0$
for
all $A \neq 0$. A state is said to be {\em normal} if and only if there is a density matrix
$\rho$ such that $\langle \varphi\mid A\rangle = \text{Tr}[\rho A]\; , \; \forall
A\in \mathfrak A$. It is called a {\em vector state} if there exists a vector $\phi \in \h$
such that $\langle \varphi \; ; \; A\rangle = \langle\phi \mid A\phi\rangle\; , \; \forall
A \in \mathfrak A$. Clearly, such a state is also normal.
A vector $\psi \in \h$ is called {\em cyclic} for the von Neumann
algebra if the set $\{A\psi \mid A \in \mathfrak A\}$ is dense in $\h$; it is said to be
{\em separating } for $\mathfrak A$ if $A\psi = B\psi\; , \; A, B \in \mathfrak A$, if and
only if $A = B$. It can then be shown that $\psi$ is cyclic for $\mathfrak A$ if and
only if it is separating for $\mathfrak A^\prime$ and vice versa. An {\em automorphism}
of the von Neumann algebra is a map $\alpha: \mathfrak A \longrightarrow \mathfrak A$
which preserves its algebraic structure. It can then be shown that $\alpha$ is norm
preserving.

The centralizer of the von Neumann algebra $\mathfrak A$, with respect to the state
$\varphi$, is the von Neumann subalgebra,
\be
  \mathfrak M_\varphi = \{B \in \mathfrak A \mid \langle \varphi \; ; \; [B,
  A]\rangle  =  \langle \varphi \; ; \;  BA - AB \rangle = 0\; , \;\;
  \forall A \in \mathfrak A \}\; .
\label{centralizer-app}
\en

 A von Neumann algebra is generated by all the unitary elements in it.
 As an example, if we take the
unitary Weyl operators (see (\ref{wigmap})) $U_\pm (x, y) = \exp [-i(xQ_\pm + yP_\pm)],\;\;
x, y \in \mathbb R$, with
$Q_\pm\; , P_\pm$ as in (\ref{quant-obs1}) and (\ref{quant-obs2}), they generate the two
von Neumann algebras $\mathfrak A_\pm$ introduced at the end of Section \ref{sec:applications}.
Similarly, a von Neumann algebra  is generated by all the projection operators contained in it.
Thus, as mentioned in Section \ref{sec:centralizer}, the centralizer algebra
$\mathfrak M_\varphi$ is generated by the projection operators $\mathbb P_i^\ell,
\;\; i =1,2,  \ldots , N$.

%----------------------------------------------------------------

\section*{Acknowledgements}
   The authors would like to acknowledge financial support from the
   Universit\`a di Palermo through Bando CORI, cap. B.U. 9.3.0002.0001.0001.
One of us (STA) would like to acknowledge a grant from the Natural Sciences and
Engineering Research Council (NSERC) of Canada.

%----------------------------------------------------------------

\end{document}